# Hierarchy in the halogen activation during surface-promoted Ullmann coupling

Néstor Merino-Díez,[a, b, c] Alejandro Pérez Paz,[d] Jingcheng Li,[b] Manuel Vilas-Varela,[e] James Lawrence,[a, c] Mohammed S. G. Mohammed,[a, c] Alejandro Berdonces-Layunta,[a, c] Ana Barragán,[c, f] J. Ignacio Pascual,[b, g] Jorge Lobo-Checa,[h] Diego Peña,[e] and Dimas G. de Oteyza*[a, c, g]

**Abstract:** Within the collection of surface-supported reactions currently accessible for the production of extended molecular nanostructures under ultra-high vacuum, Ullmann coupling has been the most successful in the controlled formation of covalent single C-C bonds. Particularly advanced control of this synthetic tool has been obtained by means of hierarchical reactivity, commonly achieved by the use of different halogen atoms that consequently display distinct activation temperatures. Here we report on the site-selective reactivity of certain carbon-halogen bonds. We use precursor molecules halogenated with bromine atoms at two non-equivalent carbon atoms and found that the Ullmann coupling occurs on Au(111) with a remarkable predilection for one of the positions. Experimental evidence is provided by means of scanning tunneling microscopy and a rationalized understanding of the observed preference is obtained from density functional theory calculations.

## Introduction

A continuous boost of computing power is at the core of current technology roadmaps. So far, the most promising approach in this pursuit involves the maximization of the number of electronic components per integrated circuit. With silicon-based technology reaching scaled-down saturation, single molecules displaying basic electronic functionalities (rectifiers, switches…) are among the most promising alternatives for the substitution of current electronic components[1]. However, an effective manufacturing of molecular electronics requires a precise control of the structure not only within the functional elements but also of their respective linkage. In this context, on-surface synthesis represents a promising platform for the implementation of molecular electronics[2]. Besides expanding the synthetic routes available for creating different forefront materials, it results in designed materials that readily feature suitable two-dimensional structure for their subsequent implementation into planar integrated circuits[3].

Aiming at their successful integration in future devices, such functional materials require mechanical robustness and high electron mobility, for which covalent bonds stand out when compared to weaker interactions[4]. Inspired by conventional wet-chemistry, numerous reactions yielding C-C bond formation have already been achieved on surfaces, including Sonogashira coupling[5], aldehyde-amine coupling[6] or Glaser coupling[7] among others[8]. Nevertheless amid this collection of C-C generating reactions, Ullmann coupling (UC), in which two aryl halides are coupled on a catalytic surface (such as the facets of commonly used coinage metals) to form a biaryl molecule, represents the most widespread one to date[9]. A milestone in the development of UC as the leading on-surface synthesis reaction scheme was set by Grill and coworkers[10], who demonstrated how the morphology and dimensionality of molecular networks can be precisely tuned by adding halogen atoms at different linking sites within the same precursor backbone. Ever since, UC has been tested on a large number of different substrates and aromatic precursors[11], highlighting its versatility for the on-surface formation of different low-dimensional nanostructures.

Advanced control of on-surface synthesis protocols has been obtained through hierarchical processes[12]. That is, controlled sequences of reactions which promote the correct step-by-step connection of molecules in the formation of complex molecular structures. With UC, this hierarchy can be achieved by functionalizing the carbon backbone with different halogens, since each of them features different energy barriers for the scission of its carbon-halogen bond[13]. In this work we show how, even when using the same halogens, a selective activation of specific C-Br bonds can be obtained by different means, namely depending on their location within the same aromatic precursor. Our combined scanning tunneling microscopy (STM) and density functional theory (DFT) results reveal that this site-selectivity is driven by the substrate and the X shape conformation of the precursor. At the

[a] N. Merino-Díez, Dr. J. Lawrence, M. S. G. Mohammed, A. Berdonces-Layunta, Dr. Dimas G. de Oteyza
Donostia International Physics Center (DIPC)
20018 San Sebastián, Spain
E-mail: d_g_oteyza@ehu.eus
[b] N. Merino-Díez, Dr. J. Li, Dr. J. I. Pascual
CIC nanoGUNE, nanoscience cooperative research center
20018 San Sebastián, Spain
[c] N. Merino-Díez, Dr. J. Lawrence, M. S. G. Mohammed, A. Berdonces-Layunta, A. Barragán, Dr. Dimas G. de Oteyza
Centro de Física de Materiales – MPC, CISC-UPV/EHU
20018 San Sebastián, Spain
[d] Dr. A. Pérez Paz
School of Physical Sciences and Nanotechnology, Yachay Tech University
100119 Urcuquí, Ecuador
[e] M. Vilas-Varela, Dr. D. Peña
CIQUS, Centro Singular de Investigación en Química Biolóxica e Materiais Moleculares
15705 Santiago de Compostela, Spain
[f] A. Barragán
Departamento de Física de Materiales, Universidad del País Vasco (UPV/EHU)
20018 San Sebastián, Spain
[g] Dr. J. I. Pascual, Dr. D. G. de Oteyza
Ikerbasque, Basque Foundation for Science
48013 Bilbao, Spain
[h] Dr. Jorge Lobo-Checa
Instituto de Ciencia de Materiales de Aragón, CSIC-Universidad de Zaragoza    50009 Zaragoza

Supporting information for this article is given via a link at the end of the document.

preferred adsorption configuration of the precursor, the halogens display different distances to the metal surface, which leads to a strongly modulated catalytic effect of the substrate for each C-Br bond.

## Results and Discussion

In this experiment we make use of 2,2',10,10'-tetrabromo-9,9'-bianthracene (TBBA, Fig. 1a, see supporting information for the synthesis and characterization of this molecule). The reactivity of similar di-brominated bianthracene (DBBA) precursors, having halogen atoms located either at 10 and 10´ or at 2 and 2´ positions, has been previously reported for different coinage metal surfaces[14]. When bromine atoms are excised preferentially at positions 2 and 2', UC governs the synthetic process and renders, after subsequent thermal cyclodehydrogenation (CDH), chiral graphene nanoribbons (chGNRs) on Au(111), Ag(111) and Cu(111) (Fig. 1e)[14]. Similarly, when bromine atoms are excised at positions 10 and 10', armchair GNRs are obtained on Au(111) and Ag(111)[14] (Fig. 1d) through UC and CDH. However, the coupling selectivity given by the halogen excision is overruled by the substrate´s catalytic effect on Cu(111). In this case, new radicals are created via dehydrogenation at $C_2$ positions, which ultimately determine the polymerization motif and yield chGNRs upon CDH[14]. The fact that the same results are obtained even with non-halogenated bianthryl precursors demonstrate that this coupling mechanism is fully independent of whether radicals are present at 10,10´ positions or not[15]. These findings highlight the complex competition in the determination of the reaction mechanism between the halogen position within the aromatic skeleton and the specific interactions with the substrate. Here, we employ TBBA precursors (Fig. 1a) on Au(111) in an attempt to specifically address this interplay.

After TBBA deposition, UC is thermally-induced at ~450 K and TBBA precursors form bianthrylene polymers. Figure 2 shows different scale STM images of this phase. As also observed with either type of DBBA reactants, the polymers appear aggregated into islands, indicating the presence of attractive intermolecular forces. These polymers are seen as a series of zigzagging lobes corresponding to the up-pointing termini of anthracene subunits. The latter arrange in an alternatingly tilted non-planar configuration to minimize the steric hindrance exerted by the neighbouring hydrogen atoms. Because this zigzag-display is shared by both armchair and chiral GNRs[14], an unambiguous determination of the polymer structure is not straightforward at this point.

**Figure 1.** Possible synthetic routes expected in this experiment from (a) TBBA precursor. (b) Debromination and $C_{10}$-$C_{10}$ Ullmann coupling (UC) followed by

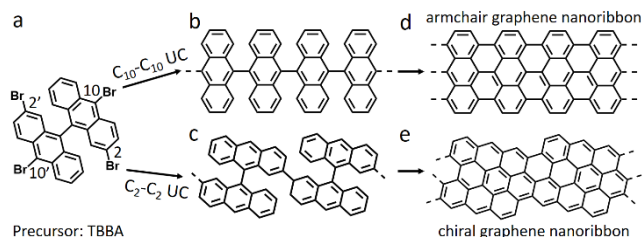

cyclodehydrogenation (CDH) leads to (d) armchair GNRs formation. (c) Debromination and $C_2$-$C_2$ UC (c) followed by CDH leads to (e) chiral GNRs, the preferred pathway here. Note that primed position x' is chemically equivalent to the corresponding unprimed position x and either would yield the same GNR.

Figure 2 also reveals the presence of a disordered network surrounding the polymeric islands and spreading over the remaining Au(111) surface. A closer look reveals that this web consists of single circular adsorbates arranged either in line or in a zigzag manner (Fig. S3). Similar networks have been observed for submonolayer coverages of halogens on Au(111)[16], as well as with other GNR precursors with a large stoichiometric halogen ratio,[17] which leads us to the conclusion that these adsorbates are bromine atoms. This kind of networks has not been observed in polymers formed from similar di-brominated precursors because the halogens are preferentially placed underneath the up-pointing anthracene ends, "hidden" from the scanning probe[18]. The observed Br network is thus assigned to the additional Br atoms as we change from dibrominated to tetrabrominated reactants, which can no longer be incorporated below the polymers. This finding underlines that both 2,2´ as well as 10,10´ positions are dehalogenated at this stage.

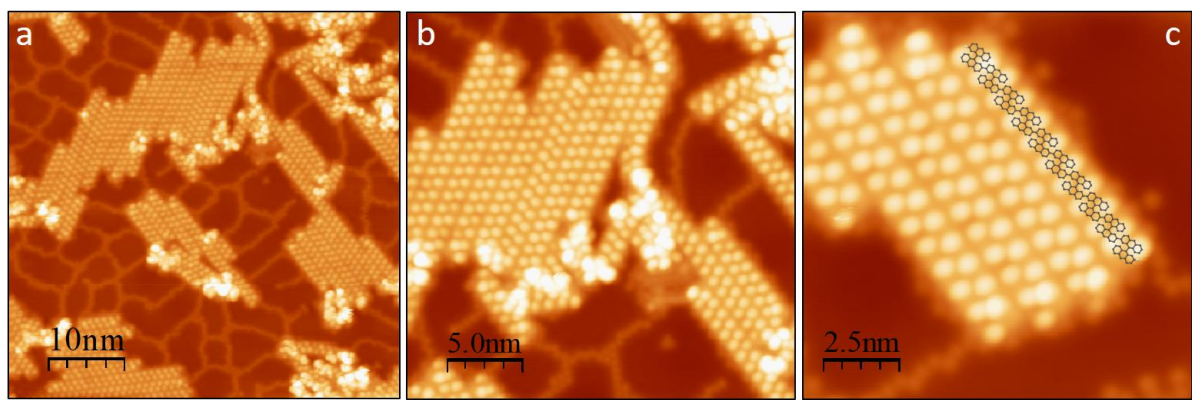

**Figure 2.** STM images of polymeric phase. (a) 50 nm² (U = 1.4 V, I = 140 pA). (b) 25 nm² (U = 1.4 V, I = 140 pA). (c) 12.5 nm² (U = 1.5 V, I = 1.0 nA) with superimposed wireframe model where only the carbon skeleton is shown, see Fig.1c?.

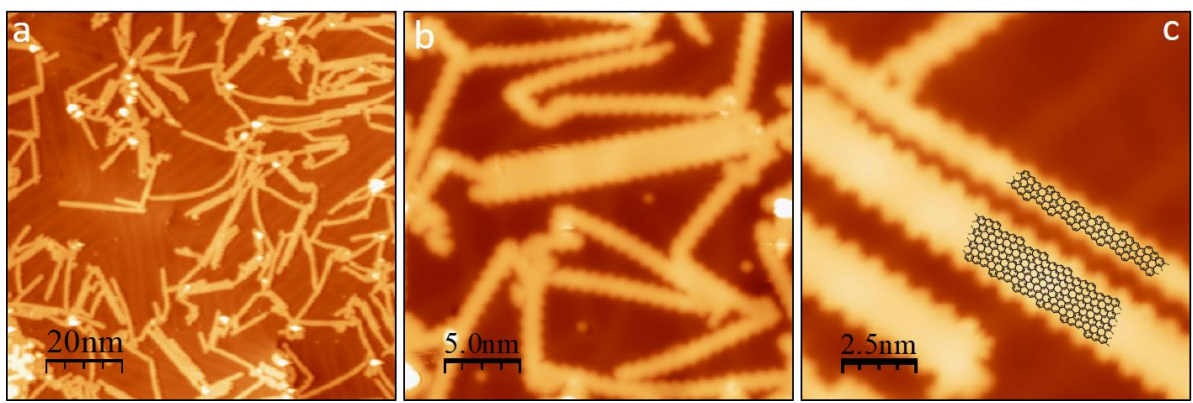

**Figure 3.** Chiral GNRs resulting from TBBA precursor. (a) 100 nm² (U = 1.1 V, I = 80 pA). (b) 25 nm² (U = 1.0 V, I = 120 pA). (c) 12.5 nm² (U = 50 mV, I = 0.5 nA) with superimposed wireframe models showing only the carbon backbone, see Fig.1e.

After polymerization, intramolecular CDH is induced by annealing the sample to higher temperatures, thereby turning the polymers into planar GNRs. At this point it is easy to identify from the lateral edge morphology that the resulting product is exclusively (3,1)-chGNRs (Fig. 3). This implies a dramatically favored UC via the 2,2' positions between the debrominated precursors, which given the well-known preferred reactivity of acenes at their central ring positions[19] is a surprising result. The UC itself being a multistep reaction, the observed preference may arise from any of the different steps. Reactant diffusion can be discarded because it would affect both halogen positions alike. A difference in the barriers associated to the bond formation between two nearby carbon radicals is also unlikely, since they are typically much lower than the barriers associated with the homolytic cleavage of the C-Br bonds[20]. Thus, the homolytic cleavage seems to be the determining step.

We have indeed proven the sequential Br activation by temperature-dependent XPS measurements. Fig. 4 (left) depicts the evolution of Br 3d core level spectra in the temperature range that displays the changes. At low temperatures, the spectra shows two different sets of Br 3d doublets (marked with green and red arrows, respectively), each associated with one of the different Br pairs. As the temperature is raised, the doublet at higher binding energies dissappears first (green arrows), with a relatively sharp transition temperature around 525 K. We thus associate this Br 3d doublet to the Br at 2,2´ positions (green arrows). Subsequently, the second doublet also fades (red arrows), although with a smoother temperature dependence. Concomitant with the dissappearance of those two core level doublets, a new one appears at a binding energy more than 2 eV lower, associated with the atomic Br adsorbed on the metal surface (blue arrows). Its intensity profile (Fig.4, right) as a function of temperature (at ~68.9 eV, see the vertical dashed blue line), reveals two distinct increases that correlate with the activation temperatures of each of the 'organic Br pairs', marked respectively with green and red horizontal lines as a guide to the eye. Finally, at around 615 K the Br 3d core level intensity dissappears, indicating the desorption of Br from the surface. Because Br desorbs preferentially as HBr,[21] this desorption temperature can be associated with the CDH temperature, at which H becomes available for Br atoms to combine with and desorb. Although the threshold temperature values extracted from the XPS and STM analyses differ, this may relate to the different chambers and temperature measurement methods used (pyrometer vs. thermocouple, respectively). However, most importantly, their combination unambiguously reveals a stepwise activation of the different Br species within the reactant.

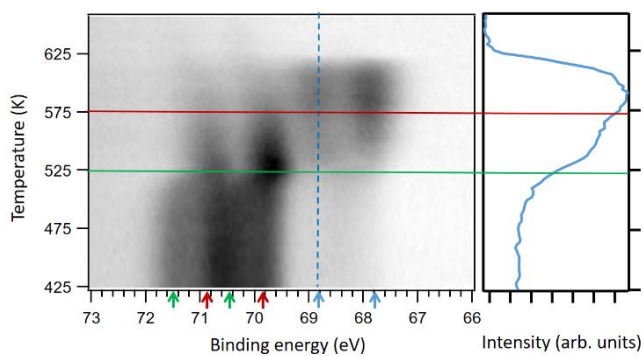

**Figure 4.** Photoemission spectra of the Br 3d core levels of precursor TBBA and their evolution as a function of sample annealing. At the right, an intensity profile at ~68.9 eV (vertical blue dashed line on the left) reveals the two distinct intensity increases with temperature. The different doublet energies for the 2,2´and 10,10´ positions, as well as for the detached Br are marked with colored arrows (green, red and blue, respectively) as a guide to the eye.

For a better understanding of the experimental findings we performed DFT calculations on this system. We studied the homolytic cleavage of C-Br bonds in TBBA and dibromoanthracene (DBA) molecules in the gas phase and found that the 10,10´ positions exhibit a slightly higher reactivity (i.e., a lower C-Br dissociation binding energy) than the 2,2´ positions with a marginal difference of 1.1 kcal/mol for both molecules, suggesting that having an additional neighboring DBA subunit (as in TBBA) has a negligible impact on the dissociation binding energy of C-Br bonds (i.e., DBA and the more sterically congested TBBA have almost identical C-Br dissociation binding energies). The preference for 2,2´ positions for TBBA on Au(111) surface must thus have a different origin. To address a possible influence of the substrate, we investigate the relative stability of two adsorption configurations in which the $C_2$-Br can be either oriented towards (Fig. 5a) or away from (Fig. 5b) the surface. Both configurations show a dihedral angle between the anthracene subunits of ~127°, consistent with the relatively low energy cost (~5 kcal/mol) needed to distort from the 90° optimum angle calculated for TBBA molecule in gas phase (Fig. S6). We find that the adsorption geometry with $C_2$-Br pointing towards the surface is 8.92 kcal/mol more favorable, making it the dominant configuration on Au(111). Although as deposited molecules are difficult to image experimentally and display notable polymorphism and disorder, the best recognizable structure indeed fits an assembly of molecules with the $C_2$-Br pointing down (Fig. S4). Interestingly, in this conformation bromine atoms located at positions 2 and 2' are closer to the surface by ~0.27 Å than those at positions 10 and 10' (Fig. 5a and Fig. S7). A difference that may be further enhanced as the temperature increases, taking into account the vibrational modes of the respective C-Br bonds (with the $C_2$-Br bonds pointing towards the surface more than $C_{10}$-Br). As the homolytic cleavage of the C-Br bonds can be catalyzed by metallic substrates[20], we associate this proximity to the substrate with an enhanced catalytic effect on the $C_2$-Br bonds, which in turn promotes the polymerization along the 2,2´ rather than along the 10,10´ positions.

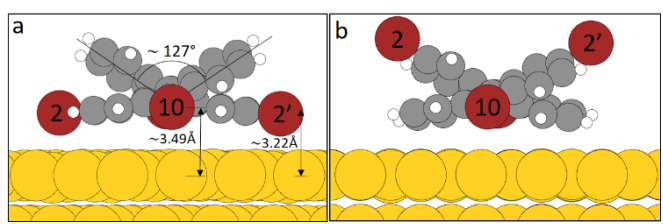

**Figure 5.** Simulated models for the adsorption of TBBA precursor on Au(111). Adsorption configuration with $C_2$-Br pointing (a) towards (the global minimum) or (b) away from the surface, 0.387 eV less stable than (a). The yellow, gray, red, and white spheres represent the Au, C, Br, and H atoms, respectively.

Lastly, it is important to note that the chiral ribbons display a variety of widths, which in turn indicates a lateral fusion of the polymers. Although the cyclodehydrogenative fusion of GNRs has been reported earlier and would lead to similar results [22], in this particular case it can be discarded. The reason for this is that while samples prepared from 2,2´-DBBA reveal no lateral fusion of the chiral ribbons after annealing to temperatures of 590 K (Fig. S5a), starting from TBBA at a similar coverage we observe wider, fused GNRs readily at 530 K (Fig. S5b). The additional halogenation thus appears to be an efficient way to increase the chiral ribbons´ widths (with its associated impact on their electronic properties[23]) with only mild annealing treatments.

Knowing that 530 K is not sufficient to drive the lateral cyclodehydrogenative fusion, the presence of radicals must be involved. It seems unlikely that GNRs would display radicals at the 10,10´ positions because they would most probably be immediately saturated by the hydrogen released in intramolecular CDH. We can thus conclude that the lateral fusion must occur in the polymeric phase, when the polymers may still be displaying the radicals at the 10,10´ positions and following the alternative coupling direction outlined in Fig. 1c. The fact that most ribbons are only one-monomer wide implies that this lateral coupling is not very effective. This may be ascribed to two different factors that may both contribute, presumably playing together. One is the chiral nature of reactants and polymers, which only allows UC between molecules with the same chirality. The deposited reactants being a racemic mixture, there is a 50% chance for two neighboring polymers to display the same chirality and thereby allow for lateral UC. In addition, the steric hindrance between the up-pointing anthryl units of neighboring polymers sharing the same chirality may interfere and further limit their coupling.

## Conclusions

In conclusion, we have evaluated the reactivity of multiple C-Br bonds located at different positions within the same aromatic precursor and observed how the relative distance of the halogens with respect to the substrate results in the selective activation at $C_2$ sites due to the closer proximity to the catalyst. This favored reactivity promotes one of the possible reaction pathways and limits the product formation to chiral GNRs. Particularly interesting is the fact that selective substrate-induced debromination at $C_2$ position overrules the well-known preferred reactivity of acenes at the central rings ($C_{10}$ positions in the case of DBBA precursor). Our results underscore the critical influence of the catalytic substrate on surface-assisted Ullmann coupling. We also demonstrate the use of surface adsorption to design new synthetic strategies that redirect the innate reactivity of aromatic molecules. Specifically, surface adsorption can afford new hierarchical synthetic routes even when carbon atoms are functionalized with the same halogen atoms.

## Experimental Section

### Experimental Methods

Samples were prepared by deposition of 2,2',10,10'-tetrabromo-9,9'-bianthracene (see supporting information for the synthesis of this molecule) precursor from Knudsen cell evaporators heated up to 475 K for sublimation. A single crystal Au(111) surface was used as substrate, prepared by standard sputtering/annealing cycles. STM samples analysis was performed at 4.3 K under ultrahigh vacuum (UHV) at pressures below $10^{-10}$ mbar. All STM images were processed by WSxM software[24].

XPS experiments were performed at the PEARL beamline at the Swiss Light Source (SLS). Samples were prepared using similar conditions to those used for STM experiments. Sample temperatures were calibrated using the readout of two optical pyrometers as a function of sample heating current. The Br 3d signal was monitored during a temperature ramp from RT to 380°C over the course of 8 hours, with each full scan taking approximately 5 minutes (99 scans total). The temperature ramp was performed by incrementally increasing the sample filament current (in direct proximity to the sample) before every scan. A photon energy of 420 eV and an analyser pass energy of 50 eV were used.

### Computational Methods

All gas phase calculations were performed using the program ORCA (version 4.0.1.2)[25]. In all calculations we included van der Waals corrections according to the Grimme's D3 approach[26] with the Becke-Johnson damping scheme (D3BJ). We used a triple-zeta quality basis set (def2-TZVP)[27] and the "TightSCF" keyword for all geometry relaxations. We performed relaxed surface scans to map out the energy profile for the rotation around the single C-C bond connecting the anthracene subunits in TBBA. We also estimated the gas-phase dissociation bond energy for the homolytic cleavage of the $C_2$-Br and $C_{10}$-Br bonds in TBBA and in 2,10-dibromoanthracene monomer. The dissociated radical species were calculated at the unrestricted level. Since we are comparing the relative strength between very similar C-Br bonds, zero point energy (ZPE) contributions were not calculated.

For gas-phase calculations, we compared different exchange-correlation functionals, including GGA (PBE)[30] and hybrid, namely, PBE0[28] and B3LYP [29]. We found that PBE provides a good compromise between accuracy and computational cost for this system and it will be the functional adopted in slab calculations. Finally, since PBE compares well with hybrid functional B3LYP in the energy profile, PBE0 energies were obtained from single point calculations on the PBE-relaxed geometries.

All slab calculations used the PBE exchange-correlation functional[30]. Since we are dealing with physisorbed molecules, we included van der Waals (vdW) corrections via the Grimme's D3 method[26]. We investigated different adsorption sites of the TBBA molecule (chemical formula $C_{28}H_{14}Br_4$) on the Au(111) surface.

An unreconstructed Au(111) surface was modeled with the coordinates derived from the experimental lattice constant of a = 4.0782 Å. The low adsorbate coverage limit was investigated using 8x8x4 slabs (256 Au atoms, 64 atoms per layer) that feature lateral adsorbate separations of ~23 Å between the centers of masses. The computational unit cell was [[L, 0.0, 0.0], [L cos(π/3), L sin(π/3), 0.0], [0.0, 0.0, 32.06365]] Å, where L=4a√2. Only the upper two Au layers are allowed to move during geometry relaxations. An ~18 Å vacuum layer and dipole corrections were used to decouple the periodic images along the normal z direction.

For slab calculations, we used the Quickstep (QS)[31,32] module of the CP2K code [version 6.1, revision number: 18464]. QS solves the electronic problem using a hybrid basis set approach that combines Gaussian and plane wave basis sets. The valence Kohn-Sham orbitals were expanded in a double-zeta quality basis set (DZVP-MOLOPT-GTH for the adsorbate, MOLOPT-DZVP-SR-GTH for Au atoms), which is specifically optimized for its use with the GTH pseudopotentials[33]. The valence electronic density was expanded using a fully converged plane wave cutoff of 600 Ry. All CP2K calculations were carried out at the converged 3x3x1 K-point sampling and employed an electronic Fermi-Dirac smearing temperature of 300 K. Geometry relaxations were stopped once the maximum ionic force on any atom fell below 1.0e-3 a.u. (0.05 eV/angstrom). Finally, we double-checked our slab CP2K calculations with the code GPAW [34] and found consistent results.

## Acknowledgements

The project leading to this publication has received funding from the European Research Council (ERC) under the European Union's Horizon 2020 research and innovation programme (grant agreement No 635919), from the Spanish Ministry of Economy, Industry and Competitiveness (MINECO, Grant Nos. MAT2016-78293-C6-R), We also acknowledge financial support from the Xunta de Galicia (Centro singular de investigación de Galicia, accreditation 2016–2019, ED431G/09) and Fondo Europeo de Desarrollo Regional (FEDER). This work used the "Imbabura" computer cluster of Yachay Tech University, which was purchased under contract No. 2017-024 (SIE-UITEY-007-2017). A.P.P. thanks D.J. Mowbray for his assistance with the computer cluster. We acknowledge the Paul Scherrer Institut, Villigen, Switzerland for provision of synchrotron radiation beamtime at PEARL beamline and would like to thank N. P. M. Bachellier and M. Muntwiller for assistance. We thank R. Fasel and R. Widmer for provision of substrate and sample holder for the synchrotron experiments.

**Keywords:** Density functional calculations • Graphene nanoribbons • Hierarchical synthesis • Scanning probe microscopy • Ullmann coupling

# Supplementary information

## 1. Synthesis and characterization of TBBA

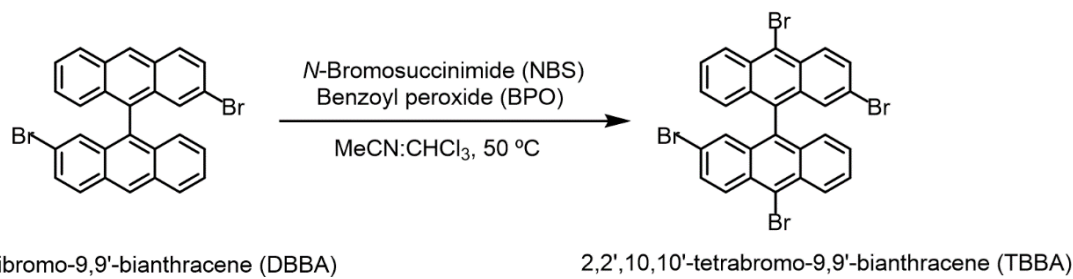

**Scheme S1**. Synthesis of 2,2',10,10'-tetrabromo-9,9'-bianthracene (TBBA).

N-Bromosuccinimide (NBS, 335 mg, 1.88 mmol) was added over a mixture of 2,2'-dibromo-9,9'-bianthracene[1] (DBBA, 64 mg, 0.12 mmol) and benzoyl peroxide (BPO, 5 mg) dissolved in MeCN:CHCl$_3$ (1:1, 8 mL). The resulting mixture was heated under argon at 50 °C for 16 h. Then, the reaction mixture was concentrated under reduced pressure and the resulting residue was purified by column chromatography (SiO$_2$, hexane/CH$_2$Cl$_2$ 9:1) to afford 2,2',10,10'-tetrabromo-9,9'-bianthracene (TBBA, 80 mg, 95% yield) as a yellow solid (mp: 362 °C).

$^1$H NMR (323 K, 500 MHz, CDCl$_3$) δ: 8.70 (d, $J$ = 8.9 Hz, 2H), 8.61 (d, $J$ = 9.4 Hz, 2H), 7.63 (dd, $J$ = 9.4, 1.9 Hz, 2H), 7.61 (ddd, $J$ = 9.0, 6.5, 1.1 Hz, 2H), 7.23 – 7.20 (m, 4H), 6.99 (d, $J$ = 8.8 Hz, 2H) ppm.

$^{13}$C NMR-DEPT (323 K, 125 MHz, CDCl$_3$) δ: 133.0 (2C), 132.9 (2C), 131.7 (2C), 131.1 (2CH), 130.9 (2C), 130.4 (2CH), 129.1 (2C), 128.6 (2CH), 128.4 (2CH), 127.9 (2CH), 127.4 (2CH), 127.1 (2CH), 125.0 (2C), 121.8 (2C) ppm.

MS (EI) $m/z$ (%): 670 (M$^+$, 100), 348 (20), 214 (16), 174 (49).

HRMS: C$_{28}$H$_{14}$Br$_4$; calculated: 665.7829, found: 665.7830

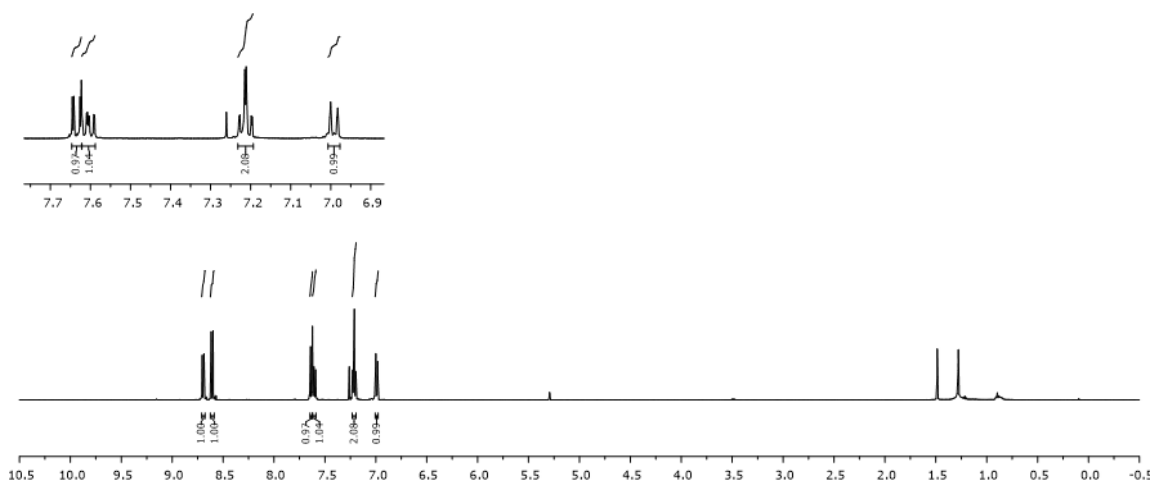

**Figure S1.** $^1$H NMR spectrum of 2,2',10,10'-tetrabromo-9,9'-bianthracene (TBBA)

---

[1] D. G. de Oteyza, A. García-Lekue, M. Vilas-Varela, N. Merino-Díez, E. Carbonell-Sanromà, M. Corso, G. Vasseur, C. Rogero, E. Guitián, J. I. Pascual, J. E. Ortega, Y. Wakayama, D. Peña, *ACS Nano* **2016**, 10, 9000-9008.

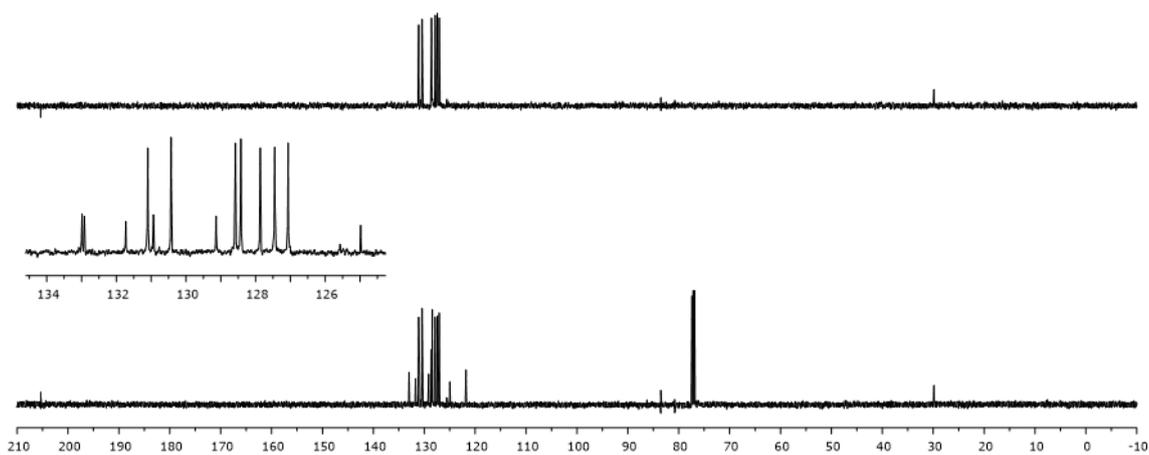

**Figure S2.** $^{13}$C NMR spectrum of 2,2',10,10'-tetrabromo-9,9'-bianthracene (TBBA)

## 2. Additional STM Images

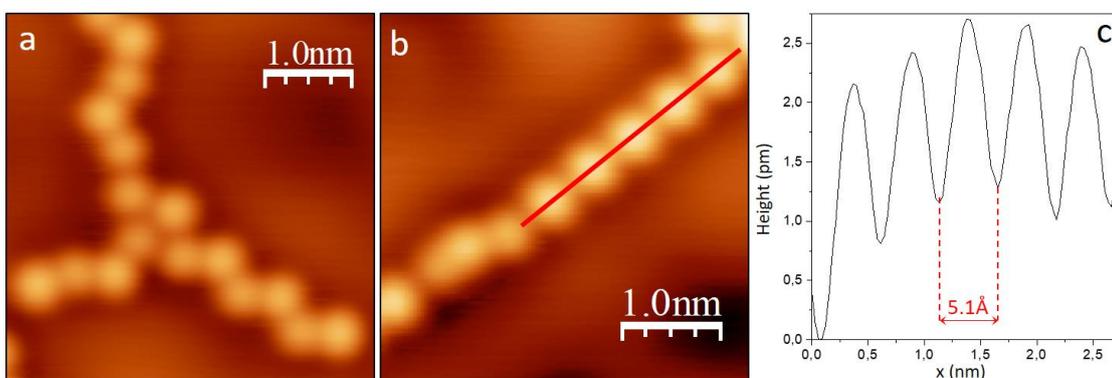

**Fig. S3.** STM images (U = 65 mV, I = 250 pA) showing bromine atoms arranged (a) in zigzag and (b) in line. (c) Line profile corresponding to the red line in (b).

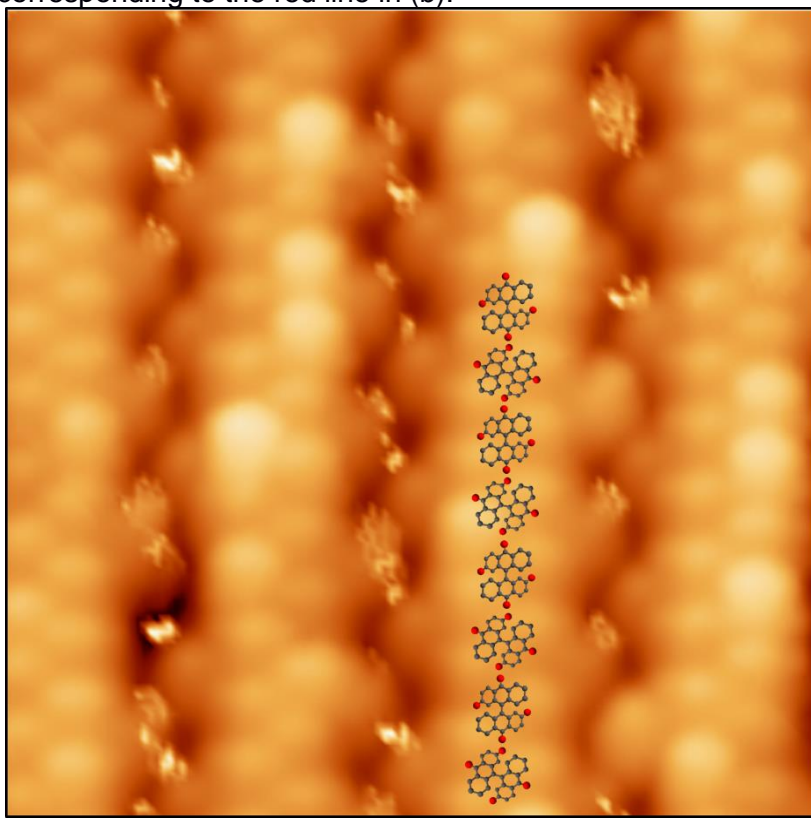

**Fig. S4. As deposited reactant molecules and tentative model of this particular ordered structure.** The molecules display alternating orientation and the structure is presumably stabilized by halogen bonding of the Br atoms along the molecular rows. Only the Br atoms at 10,10´ positions and facing out of the molecular rows are observed (every second molecule along the rows), the rest being hidden from the scanning probe by the neighboring up-pointing anthracenes.

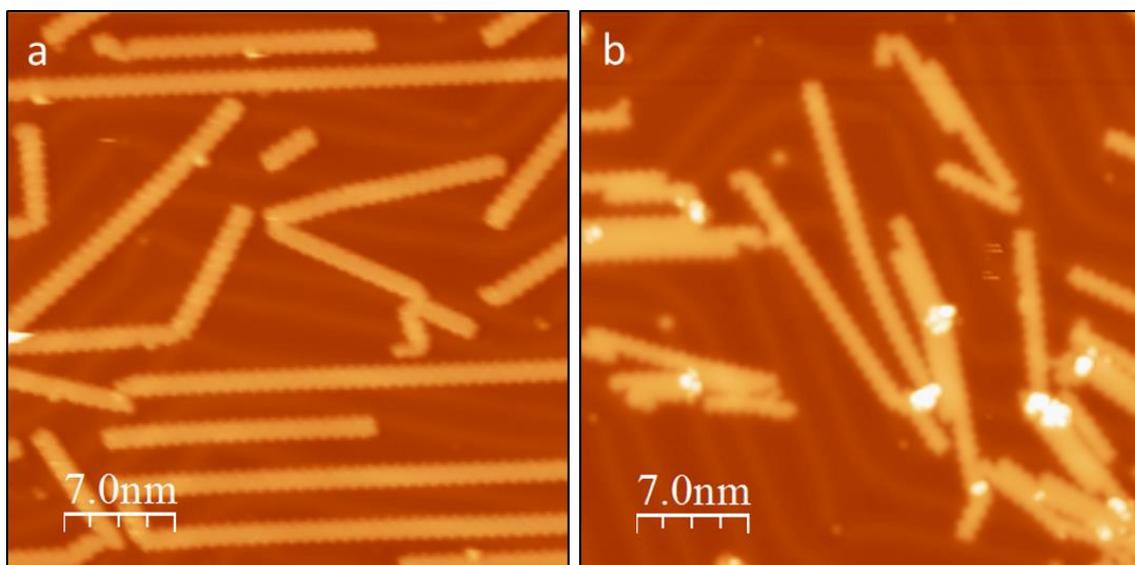

**Fig. S5.** Comparative images of samples obtained with (a) 2,2´-DBBA after annealing to 590 K (U = 0.8 V , I = 20 pA ) and with (b) TBBA after annealing to only 530 K (U = 0.8 V , I = 50 pA ).

3. Computational calculations

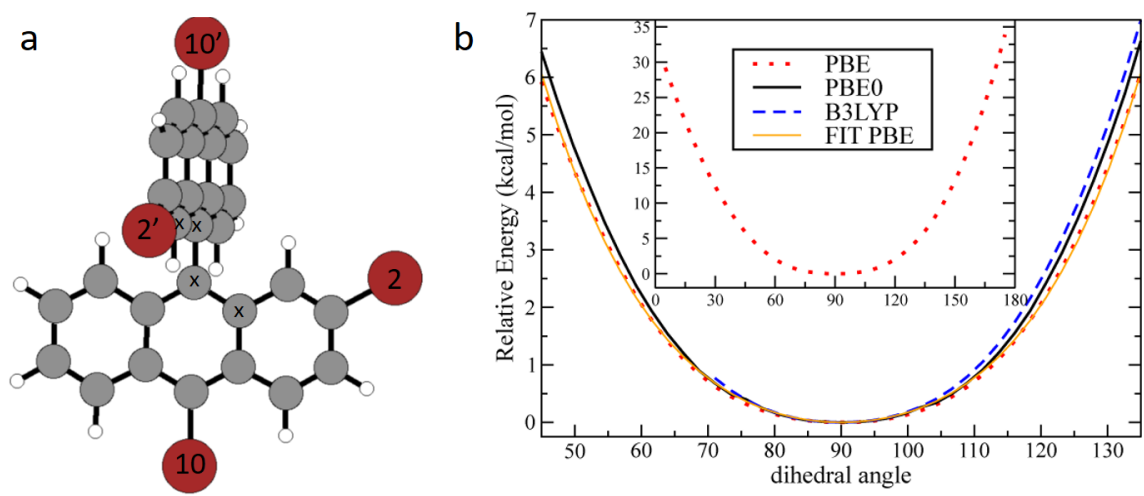

**Fig. S6. Energy cost to rotate anthracene subunits out from the 90° optimum dihedral angle in gas phase.** (a) Relaxed structure of gas-phase TBBA calculated at the PBE-D3/def2-TZVP level. The bromine positions (2, 2', 10, and 10') are indicated. This global minimum conformation features a dihedral angle (defined by the "x" symbols, 0° means eclipsing Br atoms at 2 and 2' positions) of 90°, that is, the two anthracene subunits are perpendicular. The gray, red, and white spheres represent the C, Br, and H atoms, respectively. (b) Calculated energy profile for the rotation around the single C-C bond connecting the anthracene subunits of TBBA in gas phase. Geometries were relaxed at the PBE and B3LYP level with a constraint on the dihedral angle defined in Figure S3a (0° corresponds to eclipsing Br2 and 2'). PBE0 data are single point energy calculations on the PBE-relaxed geometries. The basis set was def2-TZVP in all cases. The PBE curve can be fitted perfectly (orange line) in the range [45°,135°] by the following non-linear function: $V(a)=0.5k_h(a-90)^2+0.25k_q(a-90)^4$, where $a$ is the dihedral angle, and the fitted values are $k_h=0.00334$ and $k_q=2.621e-6$ kcal/mol for the harmonic and quartic force constants, respectively. The harmonic part alone (not shown) does already an excellent job in the range [70°, 110°] around global minimum. Outside this interval, however, the quartic potential is needed to accurately describe the full potential. Asymmetry becomes evident at the endpoints (inset shows PBE data on the full range) due to the different steric clashes between approaching phenyl rings. At these endpoints, the flanking phenyl rings distort significantly to reduce steric hindrance.

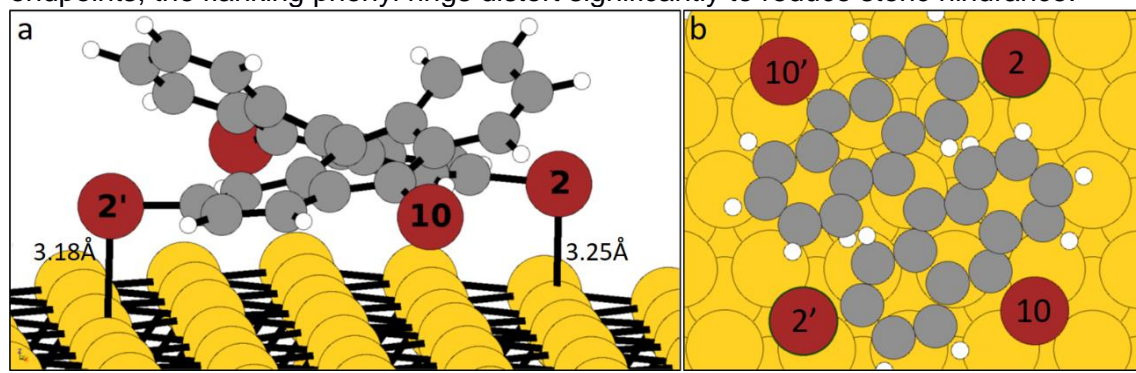

**Fig. S7. Calculated most stable TBBA adsorption configuration.** (a) Side and (b) top views of the calculated most stable adsorption conformation of TBBA on Au(111). All Br atoms are found nearly on bridge positions. The yellow, gray, red, and white spheres represent the Au, C, Br, and H atoms, respectively.

**Supplementary data 1. Relaxed XYZ coordinates from DFT calculations: a)** gas-phase TBBA structure (Fig. S3a) relaxed at the PBE-D3/def2-TZVP level using the ORCA code, (b) adsorption configuration with $C_2$-Br pointing away from the surface (Fig. 4b) and (c) detailed adsorption configuration with $C_2$-Br pointing towards the surface (global minimum, Fig. S4). The b) and c) structures were relaxed at the PBE-D3 level with the CP2K code.

<u>a) Gas-phase TBBA structure (Fig. S6a): the total energy is E=-11369.634176 Hartrees computed at the PBE-D3/def2-TZVP level with the ORCA program.</u>

```
C     1.325240    2.736104   -3.872177
C     1.270126    1.318097   -3.851679
C     0.785657    0.663786   -2.747356
C     0.328583    1.379464   -1.599203
C     0.894763    3.465797   -2.791841
C     0.383781    2.828236   -1.622673
C    -0.165815    0.700643   -0.466267
C    -0.613241    1.428727    0.655509
C    -0.070476    3.527758   -0.487973
C    -0.569214    2.877232    0.655091
C    -1.116898    0.756645    1.810171
C    -1.550115    1.471749    2.895051
C    -1.512566    2.889928    2.910912
C    -1.033353    3.563663    1.816520
H     1.712608    3.250168   -4.753101
H     1.614259    0.749284   -4.716696
H     0.739966   -0.425615   -2.727030
H     0.939000    4.554210   -2.812154
Br   -0.007616    5.441335   -0.500707
H    -1.151658   -0.332010    1.812979
Br   -2.222036    0.565169    4.428936
H    -1.862539    3.431378    3.789097
H    -1.002129    4.652732    1.826117
C    -3.677364   -2.821660   -1.797663
C    -3.623032   -1.403698   -1.820697
C    -2.498461   -0.750443   -1.383711
C    -1.362348   -1.467230   -0.899010
C    -2.608967   -3.552342   -1.339966
C    -1.419554   -2.915981   -0.876580
C    -0.210393   -0.789735   -0.448641
```

| | | | |
|---|---|---|---|
| C  | 0.898236  | -1.518985 | 0.028524 |
| C  | -0.297031 | -3.616677 | -0.394754 |
| C  | 0.864469  | -2.967414 | 0.061083 |
| C  | 2.071541  | -0.847983 | 0.488494 |
| C  | 3.142729  | -1.564032 | 0.953127 |
| C  | 3.125109  | -2.982162 | 0.992670 |
| C  | 2.012533  | -3.654927 | 0.555893 |
| H  | -4.575160 | -3.334817 | -2.145369 |
| H  | -4.478805 | -0.833911 | -2.185729 |
| H  | -2.452736 | 0.339011  | -1.398842 |
| H  | -2.655637 | -4.640709 | -1.322656 |
| Br | -0.353483 | -5.530164 | -0.357324 |
| H  | 2.099975  | 0.240620  | 0.462785 |
| Br | 4.702434  | -0.659328 | 1.565379 |
| H  | 3.992070  | -3.524394 | 1.368499 |
| H  | 1.996045  | -4.743960 | 0.585415 |

**b)** Adsorption configuration with $C_2$-Br pointing away from the surface (Fig. 5b). The calculated total energy of this configuration is -8735.9943897 Hartrees.

| | | | |
|---|---|---|---|
| Au | 0.0000000000  | 0.0000000000 | 9.2134497000 |
| Au | 2.8837228800  | 0.0000000000 | 9.2134497000 |
| Au | 5.7674457500  | 0.0000000000 | 9.2134497000 |
| Au | 8.6511686300  | 0.0000000000 | 9.2134497000 |
| Au | 11.5348915000 | 0.0000000000 | 9.2134497000 |
| Au | 14.4186143800 | 0.0000000000 | 9.2134497000 |
| Au | 17.3023372500 | 0.0000000000 | 9.2134497000 |
| Au | 20.1860601300 | 0.0000000000 | 9.2134497000 |
| Au | 1.4418614400  | 2.4973772700 | 9.2134497000 |
| Au | 4.3255843100  | 2.4973772700 | 9.2134497000 |
| Au | 7.2093071900  | 2.4973772700 | 9.2134497000 |
| Au | 10.0930300600 | 2.4973772700 | 9.2134497000 |
| Au | 12.9767529400 | 2.4973772700 | 9.2134497000 |
| Au | 15.8604758100 | 2.4973772700 | 9.2134497000 |
| Au | 18.7441986900 | 2.4973772700 | 9.2134497000 |
| Au | 21.6279215600 | 2.4973772700 | 9.2134497000 |
| Au | 2.8837228800  | 4.9947545300 | 9.2134497000 |
| Au | 5.7674457500  | 4.9947545300 | 9.2134497000 |
| Au | 8.6511686300  | 4.9947545300 | 9.2134497000 |
| Au | 11.5348915000 | 4.9947545300 | 9.2134497000 |
| Au | 14.4186143800 | 4.9947545300 | 9.2134497000 |
| Au | 17.3023372500 | 4.9947545300 | 9.2134497000 |
| Au | 20.1860601300 | 4.9947545300 | 9.2134497000 |
| Au | 23.0697830000 | 4.9947545300 | 9.2134497000 |
| Au | 4.3255843100  | 7.4921318000 | 9.2134497000 |
| Au | 7.2093071900  | 7.4921318000 | 9.2134497000 |
| Au | 10.0930300600 | 7.4921318000 | 9.2134497000 |
| Au | 12.9767529400 | 7.4921318000 | 9.2134497000 |
| Au | 15.8604758100 | 7.4921318000 | 9.2134497000 |
| Au | 18.7441986900 | 7.4921318000 | 9.2134497000 |
| Au | 21.6279215600 | 7.4921318000 | 9.2134497000 |
| Au | 24.5116444400 | 7.4921318000 | 9.2134497000 |
| Au | 5.7674457500  | 9.9895090700 | 9.2134497000 |
| Au | 8.6511686300  | 9.9895090700 | 9.2134497000 |
| Au | 11.5348915000 | 9.9895090700 | 9.2134497000 |
| Au | 14.4186143800 | 9.9895090700 | 9.2134497000 |
| Au | 17.3023372500 | 9.9895090700 | 9.2134497000 |
| Au | 20.1860601300 | 9.9895090700 | 9.2134497000 |
| Au | 23.0697830000 | 9.9895090700 | 9.2134497000 |
| Au | 25.9535058800 | 9.9895090700 | 9.2134497000 |
| Au | 7.2093071900 | 12.4868863400 | 9.2134497000 |
| Au | 10.0930300600 | 12.4868863400 | 9.2134497000 |
| Au | 12.9767529400 | 12.4868863400 | 9.2134497000 |
| Au | 15.8604758100 | 12.4868863400 | 9.2134497000 |
| Au | 18.7441986900 | 12.4868863400 | 9.2134497000 |
| Au | 21.6279215600 | 12.4868863400 | 9.2134497000 |
| Au | 24.5116444400 | 12.4868863400 | 9.2134497000 |
| Au | 27.3953673100 | 12.4868863400 | 9.2134497000 |
| Au | 8.6511686300 | 14.9842636000 | 9.2134497000 |
| Au | 11.5348915000 | 14.9842636000 | 9.2134497000 |
| Au | 14.4186143800 | 14.9842636000 | 9.2134497000 |
| Au | 17.3023372500 | 14.9842636000 | 9.2134497000 |
| Au | 20.1860601300 | 14.9842636000 | 9.2134497000 |
| Au | 23.0697830000 | 14.9842636000 | 9.2134497000 |
| Au | 25.9535058800 | 14.9842636000 | 9.2134497000 |
| Au | 28.8372287500 | 14.9842636000 | 9.2134497000 |
| Au | 10.0930300600 | 17.4816408700 | 9.2134497000 |
| Au | 12.9767529400 | 17.4816408700 | 9.2134497000 |
| Au | 15.8604758100 | 17.4816408700 | 9.2134497000 |
| Au | 18.7441986900 | 17.4816408700 | 9.2134497000 |
| Au | 21.6279215600 | 17.4816408700 | 9.2134497000 |
| Au | 24.5116444400 | 17.4816408700 | 9.2134497000 |
| Au | 27.3953673100 | 17.4816408700 | 9.2134497000 |
| Au | 30.2790901900 | 17.4816408700 | 9.2134497000 |
| Au | 1.4418614400 | 0.8324590900 | 11.5679995600 |
| Au | 4.3255843100 | 0.8324590900 | 11.5679995600 |
| Au | 7.2093071900 | 0.8324590900 | 11.5679995600 |
| Au | 10.0930300600 | 0.8324590900 | 11.5679995600 |
| Au | 12.9767529400 | 0.8324590900 | 11.5679995600 |
| Au | 15.8604758100 | 0.8324590900 | 11.5679995600 |
| Au | 18.7441986900 | 0.8324590900 | 11.5679995600 |
| Au | 21.6279215600 | 0.8324590900 | 11.5679995600 |
| Au | 2.8837228800 | 3.3298363600 | 11.5679995600 |
| Au | 5.7674457500 | 3.3298363600 | 11.5679995600 |
| Au | 8.6511686300 | 3.3298363600 | 11.5679995600 |
| Au | 11.5348915000 | 3.3298363600 | 11.5679995600 |
| Au | 14.4186143800 | 3.3298363600 | 11.5679995600 |
| Au | 17.3023372500 | 3.3298363600 | 11.5679995600 |
| Au | 20.1860601300 | 3.3298363600 | 11.5679995600 |
| Au | 23.0697830000 | 3.3298363600 | 11.5679995600 |
| Au | 4.3255843100 | 5.8272136200 | 11.5679995600 |
| Au | 7.2093071900 | 5.8272136200 | 11.5679995600 |
| Au | 10.0930300600 | 5.8272136200 | 11.5679995600 |
| Au | 12.9767529400 | 5.8272136200 | 11.5679995600 |
| Au | 15.8604758100 | 5.8272136200 | 11.5679995600 |
| Au | 18.7441986900 | 5.8272136200 | 11.5679995600 |
| Au | 21.6279215600 | 5.8272136200 | 11.5679995600 |
| Au | 24.5116444400 | 5.8272136200 | 11.5679995600 |
| Au | 5.7674457500 | 8.3245908900 | 11.5679995600 |
| Au | 8.6511686300 | 8.3245908900 | 11.5679995600 |
| Au | 11.5348915000 | 8.3245908900 | 11.5679995600 |
| Au | 14.4186143800 | 8.3245908900 | 11.5679995600 |
| Au | 17.3023372500 | 8.3245908900 | 11.5679995600 |
| Au | 20.1860601300 | 8.3245908900 | 11.5679995600 |
| Au | 23.0697830000 | 8.3245908900 | 11.5679995600 |

```
Au    25.9535058800     8.3245908900    11.5679995600
Au     7.2093071900    10.8219681600    11.5679995600
Au    10.0930300600    10.8219681600    11.5679995600
Au    12.9767529400    10.8219681600    11.5679995600
Au    15.8604758100    10.8219681600    11.5679995600
Au    18.7441986900    10.8219681600    11.5679995600
Au    21.6279215600    10.8219681600    11.5679995600
Au    24.5116444400    10.8219681600    11.5679995600
Au    27.3953673100    10.8219681600    11.5679995600
Au     8.6511686300    13.3193454300    11.5679995600
Au    11.5348915000    13.3193454300    11.5679995600
Au    14.4186143800    13.3193454300    11.5679995600
Au    17.3023372500    13.3193454300    11.5679995600
Au    20.1860601300    13.3193454300    11.5679995600
Au    23.0697830000    13.3193454300    11.5679995600
Au    25.9535058800    13.3193454300    11.5679995600
Au    28.8372287500    13.3193454300    11.5679995600
Au    10.0930300600    15.8167226900    11.5679995600
Au    12.9767529400    15.8167226900    11.5679995600
Au    15.8604758100    15.8167226900    11.5679995600
Au    18.7441986900    15.8167226900    11.5679995600
Au    21.6279215600    15.8167226900    11.5679995600
Au    24.5116444400    15.8167226900    11.5679995600
Au    27.3953673100    15.8167226900    11.5679995600
Au    30.2790901900    15.8167226900    11.5679995600
Au    11.5348915000    18.3140999600    11.5679995600
Au    14.4186143800    18.3140999600    11.5679995600
Au    17.3023372500    18.3140999600    11.5679995600
Au    20.1860601300    18.3140999600    11.5679995600
Au    23.0697830000    18.3140999600    11.5679995600
Au    25.9535058800    18.3140999600    11.5679995600
Au    28.8372287500    18.3140999600    11.5679995600
Au    31.7209516300    18.3140999600    11.5679995600
Au    -0.0076251868     1.6597136165    14.0337474118
Au     2.8772273929     1.6576271885    14.0334764794
Au     5.7610766739     1.6564088413    14.0353806378
Au     8.6454657610     1.6568538066    14.0359880371
Au    11.5306313861     1.6588973485    14.0352924969
Au    14.4178230162     1.6596395564    14.0414786722
Au    17.2988832214     1.6616067113    14.0428582843
Au    20.1785538359     1.6620341212    14.0360487157
Au     1.4350252443     4.1548598556    14.0340759500
Au     4.3179970953     4.1556939437    14.0345557105
Au     7.1976180641     4.1521279497    14.0346506393
Au    10.0811304729     4.1515838106    14.0223393321
Au    12.9796153101     4.1509987328    14.0231642474
Au    15.8653423200     4.1540598392    14.0433711647
Au    18.7390234016     4.1589569211    14.0496051027
Au    21.6191628110     4.1572019870    14.0396351721
Au     2.8786046338     6.6537767853    14.0350701985
Au     5.7590109149     6.6551119482    14.0367330794
Au     8.6343697018     6.6499130299    14.0119123882
Au    11.5323980929     6.6568681993    14.0134477446
Au    14.4170850238     6.6539068505    14.0364511762
Au    17.2929161386     6.6479363802    14.0400431227
Au    20.1824571836     6.6492666203    14.0354213918
Au    23.0675564673     6.6490629773    14.0353831030
```

```
Au      4.3225171322     9.1529111068    14.0354266937
Au      7.2048513015     9.1549532425    14.0366694251
Au     10.0864342908     9.1678352409    14.0162471608
Au     12.9697967478     9.1644843267    14.0318560988
Au     15.8435899392     9.1475576736    14.0338347451
Au     18.7315280582     9.1414440706    14.0121523324
Au     21.6349763043     9.1425088738    14.0126078987
Au     24.5118693467     9.1509203532    14.0363197226
Au      5.7634908080    11.6510407908    14.0342515367
Au      8.6465854405    11.6570023354    14.0398850173
Au     11.5300611883    11.6650085054    14.0370835876
Au     14.4065223525    11.6539548532    14.0392444801
Au     17.2932352536    11.6481660763    14.0298269887
Au     20.1823317763    11.6549035150    14.0136187043
Au     23.0679360156    11.6485153942    14.0318379124
Au     25.9495202797    11.6498959981    14.0350919598
Au      7.2037985517    14.1497467708    14.0340635507
Au     10.0886997019    14.1508258143    14.0403964988
Au     12.9726489631    14.1483260569    14.0493593890
Au     15.8542819757    14.1495872054    14.0387316896
Au     18.7373203014    14.1551159630    14.0298035396
Au     21.6234852719    14.1520657534    14.0341969631
Au     24.5054720206    14.1472045950    14.0358984214
Au     27.3895568008    14.1475818912    14.0327707322
Au      8.6455428276    16.6457776492    14.0340079443
Au     11.5303020481    16.6433905387    14.0374683188
Au     14.4134625851    16.6433171405    14.0398304936
Au     17.2959673683    16.6461670825    14.0370707039
Au     20.1810614495    16.6465922448    14.0357819640
Au     23.0651333838    16.6460389158    14.0355532327
Au     25.9476138542    16.6456610471    14.0336085603
Au     28.8308881303    16.6461692835    14.0329905755
Au     10.0863419844    19.1420446457    14.0336304319
Au     12.9709192393    19.1392904993    14.0340665637
Au     15.8544073894    19.1396861954    14.0343472187
Au     18.7393308408    19.1396026767    14.0360822685
Au     21.6246297506    19.1423953016    14.0361518084
Au     24.5080352973    19.1439993043    14.0355640652
Au     27.3902466592    19.1434433952    14.0366923652
Au     30.2722179281    19.1441251766    14.0344519359
Au     -0.0114168830    -0.0148259933    16.5706700700
Au      2.8724554433    -0.0158904883    16.5732296388
Au      5.7544355792    -0.0152614567    16.5747009537
Au      8.6363727894    -0.0142382916    16.5763917668
Au     11.5201966698    -0.0123979007    16.5808829128
Au     14.4042116101    -0.0111637589    16.5889327495
Au     17.2871685627    -0.0094674925    16.5861421270
Au     20.1718727072    -0.0111553098    16.5738683657
Au      1.4297774754     2.4822762616    16.5719987414
Au      4.3135891472     2.4836072744    16.5720986722
Au      7.1972829380     2.4854610197    16.5777657648
Au     10.0785958162     2.4829064863    16.5855126770
Au     12.9627965498     2.4800925561    16.5835473525
Au     15.8455632811     2.4836025738    16.5906028427
Au     18.7274933893     2.4853894478    16.5950351585
Au     21.6145562064     2.4813720884    16.5777566062
Au      2.8691948289     4.9815377694    16.5729872754
```

```
Au    5.7532554342    4.9830256043   16.5763717287
Au    8.6363913954    4.9852388153   16.5703483156
Au   11.5206911719    4.9753361068   16.4947003149
Au   14.4068689099    4.9781971786   16.5518549531
Au   17.2868346191    4.9770788892   16.6326356630
Au   20.1719312264    4.9713290091   16.6070807938
Au   23.0556100021    4.9781936214   16.5789925969
Au    4.3069975274    7.4809390788   16.5765080527
Au    7.1857878419    7.4837137124   16.5806878877
Au   10.0555754196    7.4942798174   16.4720243482
Au   12.9687499012    7.5046727671   16.5094887801
Au   15.8453009352    7.4652078503   16.6118722938
Au   18.7289017916    7.4521032340   16.5670667873
Au   21.6121606095    7.4695594704   16.5814267434
Au   24.4937692145    7.4785865997   16.5769514386
Au    5.7496538299    9.9816396810   16.5751584810
Au    8.6305845058    9.9892279485   16.5858568705
Au   11.5137008532    9.9993688042   16.5625704476
Au   14.4081631438    9.9880691884   16.6106669539
Au   17.2764824970    9.9481522774   16.5180517994
Au   20.1803741551    9.9637713659   16.4572636850
Au   23.0553023951    9.9752057194   16.5793843806
Au   25.9366281904    9.9783380013   16.5772412705
Au    7.1930223410   12.4811818153   16.5768620396
Au   10.0775395105   12.4847274192   16.6009586456
Au   12.9625798412   12.4737440563   16.6295776779
Au   15.8416493137   12.4670263025   16.5582007859
Au   18.7252186847   12.4783712563   16.5185732082
Au   21.6153366707   12.4749787952   16.5635667419
Au   24.4961053503   12.4750521972   16.5767581686
Au   27.3792484469   12.4779775406   16.5728454737
Au    8.6370481773   14.9768476052   16.5771972861
Au   11.5225407327   14.9704345277   16.5958872723
Au   14.4002498843   14.9678957977   16.5833240290
Au   17.2859259775   14.9705798200   16.5862216836
Au   20.1710096216   14.9710940087   16.5890040511
Au   23.0534806965   14.9726646456   16.5795803671
Au   25.9376938692   14.9745522245   16.5716168322
Au   28.8217747147   14.9766079244   16.5715170329
Au   10.0815204490   17.4695748848   16.5732698077
Au   12.9653116437   17.4660120288   16.5837307512
Au   15.8469826371   17.4641267338   16.5857278182
Au   18.7284901471   17.4649633198   16.5839888550
Au   21.6116532115   17.4687363286   16.5764067487
Au   24.4961611227   17.4712823349   16.5733672266
Au   27.3802393801   17.4729349983   16.5736939953
Au   30.2649355650   17.4733092707   16.5710053328
Br   15.6307791729    3.2893202617   19.9819460921
Br   10.8129891889    9.5497096251   23.2819903412
Br   19.8037738754    8.0855152031   23.5618214417
Br   15.2656383681   14.2295525974   19.9041825896
C    15.3516010593   12.3385686200   20.2761281512
C    14.3289014486   10.2731065888   21.1027413831
C    15.4079124712    9.5052283426   20.5779148245
C    16.5518456860   11.6530660748   20.0073487277
C    12.0171915237   11.7430151362   21.8659423550
C    16.5304460761   10.1998756278   20.0440832793
```

| | | | |
|---|---|---|---|
| C | 17.7138772844 | 9.5297727909 | 19.5926706855 |
| C | 18.9014324674 | 10.2068599359 | 19.4150522671 |
| C | 18.9490044539 | 11.6194154393 | 19.5272024363 |
| C | 13.3028882514 | 9.6732394414 | 21.8932648209 |
| C | 12.1888404913 | 10.3889936648 | 22.2528429683 |
| C | 14.2301056648 | 11.6946027595 | 20.8268822118 |
| C | 17.7860799119 | 12.3230063518 | 19.7466371954 |
| C | 11.8958522960 | 5.8126078587 | 19.5054672591 |
| C | 15.4836796571 | 5.1746361832 | 20.3559024489 |
| C | 11.9238980661 | 7.2230604909 | 19.3558637673 |
| C | 13.0950763933 | 7.9270138844 | 19.5492796342 |
| C | 14.2767753580 | 7.2875997745 | 20.0458599467 |
| C | 13.0665913559 | 5.1355160804 | 19.7583863982 |
| C | 14.2813851425 | 5.8336455880 | 20.0362600540 |
| C | 15.3688368179 | 8.0112301296 | 20.6024828550 |
| C | 16.4300908571 | 7.2699393642 | 21.1985764810 |
| C | 13.0260700946 | 12.3747657585 | 21.1828522362 |
| C | 17.4052841051 | 7.9023861079 | 22.0263449587 |
| C | 18.5020002505 | 7.2076119474 | 22.4708523176 |
| C | 18.7076497530 | 5.8457944458 | 22.1312329603 |
| C | 17.7497806630 | 5.1873113231 | 21.4014863185 |
| C | 16.5623736137 | 5.8449531294 | 20.9586207648 |
| H | 17.8035667669 | 13.4110960109 | 19.7718845267 |
| H | 19.8976454945 | 12.1449986677 | 19.4188180954 |
| H | 19.8093148712 | 9.6462854651 | 19.1858836095 |
| H | 17.6999995489 | 8.4499796487 | 19.4727177262 |
| H | 13.4055341371 | 8.6325207759 | 22.1929252764 |
| H | 17.8778820747 | 4.1302415424 | 21.1739986491 |
| H | 19.6022704678 | 5.3266797196 | 22.4721831113 |
| H | 17.2759136864 | 8.9495869186 | 22.2910725945 |
| H | 13.0688041044 | 4.0479249992 | 19.8072215520 |
| H | 13.0902987264 | 9.0042750494 | 19.4096493317 |
| H | 11.0109814306 | 7.7629060153 | 19.1007067734 |
| H | 10.9561372210 | 5.2698974006 | 19.4001445137 |
| H | 11.1072944620 | 12.2773966820 | 22.1353659104 |
| H | 12.9226706013 | 13.4265207741 | 20.9209036062 |

**c)** Detailed adsorption configuration with C$_2$-Br pointing towards from the surface (the global adsorption minimum, Fig. S7). The calculated total energy of this configuration is -8736.008610 Hartrees.
Au 0.00000000 0.00000000 9.21344970
Au 2.88372288 0.00000000 9.21344970
Au 5.76744575 0.00000000 9.21344970
Au 8.65116863 0.00000000 9.21344970
Au 11.53489150 0.00000000 9.21344970
Au 14.41861438 0.00000000 9.21344970
Au 17.30233725 0.00000000 9.21344970
Au 20.18606013 0.00000000 9.21344970
Au 1.44186144 2.49737727 9.21344970
Au 4.32558431 2.49737727 9.21344970
Au 7.20930719 2.49737727 9.21344970
Au 10.09303006 2.49737727 9.21344970
Au 12.97675294 2.49737727 9.21344970
Au 15.86047581 2.49737727 9.21344970
Au 18.74419869 2.49737727 9.21344970
Au 21.62792156 2.49737727 9.21344970
Au 2.88372288 4.99475453 9.21344970

```
Au 5.76744575 4.99475453 9.21344970
Au 8.65116863 4.99475453 9.21344970
Au 11.53489150 4.99475453 9.21344970
Au 14.41861438 4.99475453 9.21344970
Au 17.30233725 4.99475453 9.21344970
Au 20.18606013 4.99475453 9.21344970
Au 23.06978300 4.99475453 9.21344970
Au 4.32558431 7.49213180 9.21344970
Au 7.20930719 7.49213180 9.21344970
Au 10.09303006 7.49213180 9.21344970
Au 12.97675294 7.49213180 9.21344970
Au 15.86047581 7.49213180 9.21344970
Au 18.74419869 7.49213180 9.21344970
Au 21.62792156 7.49213180 9.21344970
Au 24.51164444 7.49213180 9.21344970
Au 5.76744575 9.98950907 9.21344970
Au 8.65116863 9.98950907 9.21344970
Au 11.53489150 9.98950907 9.21344970
Au 14.41861438 9.98950907 9.21344970
Au 17.30233725 9.98950907 9.21344970
Au 20.18606013 9.98950907 9.21344970
Au 23.06978300 9.98950907 9.21344970
Au 25.95350588 9.98950907 9.21344970
Au 7.20930719 12.48688634 9.21344970
Au 10.09303006 12.48688634 9.21344970
Au 12.97675294 12.48688634 9.21344970
Au 15.86047581 12.48688634 9.21344970
Au 18.74419869 12.48688634 9.21344970
Au 21.62792156 12.48688634 9.21344970
Au 24.51164444 12.48688634 9.21344970
Au 27.39536731 12.48688634 9.21344970
Au 8.65116863 14.98426360 9.21344970
Au 11.53489150 14.98426360 9.21344970
Au 14.41861438 14.98426360 9.21344970
Au 17.30233725 14.98426360 9.21344970
Au 20.18606013 14.98426360 9.21344970
Au 23.06978300 14.98426360 9.21344970
Au 25.95350588 14.98426360 9.21344970
Au 28.83722875 14.98426360 9.21344970
Au 10.09303006 17.48164087 9.21344970
Au 12.97675294 17.48164087 9.21344970
Au 15.86047581 17.48164087 9.21344970
Au 18.74419869 17.48164087 9.21344970
Au 21.62792156 17.48164087 9.21344970
Au 24.51164444 17.48164087 9.21344970
Au 27.39536731 17.48164087 9.21344970
Au 30.27909019 17.48164087 9.21344970
Au 1.44186144 0.83245909 11.56799956
Au 4.32558431 0.83245909 11.56799956
Au 7.20930719 0.83245909 11.56799956
Au 10.09303006 0.83245909 11.56799956
Au 12.97675294 0.83245909 11.56799956
Au 15.86047581 0.83245909 11.56799956
Au 18.74419869 0.83245909 11.56799956
Au 21.62792156 0.83245909 11.56799956
Au 2.88372288 3.32983636 11.56799956
Au 5.76744575 3.32983636 11.56799956
```

```
Au 8.65116863 3.32983636 11.56799956
Au 11.53489150 3.32983636 11.56799956
Au 14.41861438 3.32983636 11.56799956
Au 17.30233725 3.32983636 11.56799956
Au 20.18606013 3.32983636 11.56799956
Au 23.06978300 3.32983636 11.56799956
Au 4.32558431 5.82721362 11.56799956
Au 7.20930719 5.82721362 11.56799956
Au 10.09303006 5.82721362 11.56799956
Au 12.97675294 5.82721362 11.56799956
Au 15.86047581 5.82721362 11.56799956
Au 18.74419869 5.82721362 11.56799956
Au 21.62792156 5.82721362 11.56799956
Au 24.51164444 5.82721362 11.56799956
Au 5.76744575 8.32459089 11.56799956
Au 8.65116863 8.32459089 11.56799956
Au 11.53489150 8.32459089 11.56799956
Au 14.41861438 8.32459089 11.56799956
Au 17.30233725 8.32459089 11.56799956
Au 20.18606013 8.32459089 11.56799956
Au 23.06978300 8.32459089 11.56799956
Au 25.95350588 8.32459089 11.56799956
Au 7.20930719 10.82196816 11.56799956
Au 10.09303006 10.82196816 11.56799956
Au 12.97675294 10.82196816 11.56799956
Au 15.86047581 10.82196816 11.56799956
Au 18.74419869 10.82196816 11.56799956
Au 21.62792156 10.82196816 11.56799956
Au 24.51164444 10.82196816 11.56799956
Au 27.39536731 10.82196816 11.56799956
Au 8.65116863 13.31934543 11.56799956
Au 11.53489150 13.31934543 11.56799956
Au 14.41861438 13.31934543 11.56799956
Au 17.30233725 13.31934543 11.56799956
Au 20.18606013 13.31934543 11.56799956
Au 23.06978300 13.31934543 11.56799956
Au 25.95350588 13.31934543 11.56799956
Au 28.83722875 13.31934543 11.56799956
Au 10.09303006 15.81672269 11.56799956
Au 12.97675294 15.81672269 11.56799956
Au 15.86047581 15.81672269 11.56799956
Au 18.74419869 15.81672269 11.56799956
Au 21.62792156 15.81672269 11.56799956
Au 24.51164444 15.81672269 11.56799956
Au 27.39536731 15.81672269 11.56799956
Au 30.27909019 15.81672269 11.56799956
Au 11.53489150 18.31409996 11.56799956
Au 14.41861438 18.31409996 11.56799956
Au 17.30233725 18.31409996 11.56799956
Au 20.18606013 18.31409996 11.56799956
Au 23.06978300 18.31409996 11.56799956
Au 25.95350588 18.31409996 11.56799956
Au 28.83722875 18.31409996 11.56799956
Au 31.72095163 18.31409996 11.56799956
Au -0.00057997 1.66204397 14.03537198
Au 2.88233974 1.66322702 14.03518149
Au 5.76607917 1.66236858 14.03663819
```

```
Au 8.65014224 1.65996611 14.03421039
Au 11.53727716 1.65970157 14.03356056
Au 14.42479684 1.65945065 14.03561208
Au 17.30697360 1.66104494 14.03826702
Au 20.18727040 1.66230420 14.03619625
Au 1.44152995 4.16066928 14.03378529
Au 4.32357680 4.16112111 14.03478671
Au 7.20151327 4.15731926 14.02617459
Au 10.08486357 4.15534052 14.01537384
Au 12.98418836 4.15513800 14.01558371
Au 15.87170843 4.15572649 14.02532673
Au 18.75101742 4.15949555 14.03411634
Au 21.63115478 4.15958947 14.03049971
Au 2.88384596 6.65816158 14.03313961
Au 5.76785009 6.65927535 14.03505019
Au 8.64820590 6.66033561 14.02722344
Au 11.53504590 6.66627033 14.01570537
Au 14.42746491 6.66248866 14.02115848
Au 17.30871119 6.66050364 14.03801281
Au 20.18747918 6.65840677 14.04080548
Au 23.06935049 6.65738267 14.03482470
Au 4.32830385 9.15519305 14.03469747
Au 7.21222147 9.15780861 14.04115621
Au 10.09307336 9.16758395 14.03808687
Au 12.97551765 9.17096277 14.03240654
Au 15.86122650 9.16214902 14.04659566
Au 18.74497563 9.15395639 14.04914500
Au 21.62783767 9.15160889 14.04189094
Au 24.51254195 9.15353963 14.03456738
Au 5.76829926 11.65208837 14.03585935
Au 8.64594856 11.65316896 14.04427591
Au 11.52506749 11.65371327 14.04299582
Au 14.41051272 11.64503150 14.02281947
Au 17.30898279 11.64169567 14.02391866
Au 20.19803689 11.64227797 14.03054944
Au 23.07536577 11.64831496 14.03600106
Au 25.95644253 11.65184210 14.03356615
Au 7.20817563 14.14905056 14.03435085
Au 10.08825001 14.14943446 14.03791371
Au 12.97025456 14.14794764 14.02900024
Au 15.85808269 14.14637066 14.01508807
Au 18.75458332 14.14470619 14.01491192
Au 21.63705520 14.14554709 14.02860461
Au 24.51495407 14.14958725 14.03468764
Au 27.39762659 14.14981421 14.03379546
Au 8.65245489 16.64833031 14.03410774
Au 11.53401367 16.64978274 14.03638206
Au 14.41656443 16.65345937 14.03031951
Au 17.30423277 16.65813572 14.01926090
Au 20.18934059 16.65050970 14.03012427
Au 23.07190110 16.64721670 14.03455299
Au 25.95545638 16.64755580 14.03305371
Au 28.83937811 16.64680601 14.03374894
Au 10.09382714 19.14375618 14.03622734
Au 12.97635702 19.14481824 14.03718328
Au 15.85931168 19.14847479 14.03511141
Au 18.74582063 19.14838552 14.03515584
```

```
Au 21.63043807 19.14440512 14.03751950
Au 24.51420076 19.14491198 14.03633262
Au 27.39774617 19.14407995 14.03586864
Au 30.28060208 19.14331931 14.03591740
Au 0.00134099 -0.00697120 16.57723661
Au 2.88397422 -0.00713511 16.57993787
Au 5.76774506 -0.00645582 16.57987169
Au 8.65246879 -0.00473443 16.58160994
Au 11.53873145 -0.00334546 16.57982299
Au 14.42360678 -0.00254148 16.58149310
Au 17.30676991 -0.00181007 16.57806010
Au 20.18914659 -0.00403806 16.57505705
Au 1.44320598 2.49119846 16.57324012
Au 4.32668382 2.49120324 16.57390127
Au 7.20986933 2.49221008 16.57773332
Au 10.09213753 2.48759368 16.57220861
Au 12.98358067 2.48926799 16.57792412
Au 15.86571078 2.49344128 16.59291710
Au 18.74750460 2.49567471 16.58945716
Au 21.63039320 2.49206977 16.57827003
Au 2.88475343 4.98950643 16.57360120
Au 5.76536299 4.99083011 16.57781785
Au 8.64157851 4.99224737 16.54346521
Au 11.53531434 4.97762195 16.46219094
Au 14.43208074 4.99264247 16.51377312
Au 17.31015998 4.99001696 16.56057427
Au 20.19490178 4.98794669 16.57315299
Au 23.07317068 4.98939147 16.57756141
Au 4.32681631 7.48911661 16.57304286
Au 7.20697547 7.49187154 16.58443589
Au 10.08444442 7.50952475 16.52670633
Au 12.98852905 7.51468906 16.49678964
Au 15.87641175 7.49917860 16.56444650
Au 18.75771147 7.49299768 16.58360409
Au 21.63537794 7.48582776 16.59133005
Au 24.51470933 7.48743312 16.57406031
Au 5.77069876 9.98800321 16.58007050
Au 8.65145512 9.99176064 16.61845011
Au 11.53751342 9.98973857 16.64535137
Au 14.42319758 9.98390212 16.63978221
Au 17.30729126 9.97523939 16.64338840
Au 20.19174120 9.97828284 16.61579179
Au 23.07153131 9.98444692 16.58036407
Au 25.95590437 9.98609396 16.57144253
Au 7.20959786 12.48555150 16.59098272
Au 10.08826276 12.47468306 16.57912238
Au 12.96965820 12.46926734 16.55726819
Au 15.85732064 12.45599807 16.49084379
Au 18.75502577 12.46016480 16.51868944
Au 21.63249448 12.48113203 16.58158334
Au 24.51465458 12.48264498 16.57433257
Au 27.39798839 12.48400582 16.57496856
Au 8.64736518 14.98207104 16.57366267
Au 11.53386957 14.98007361 16.56098773
Au 14.41164140 14.97709088 16.52506552
Au 17.30532391 14.99305720 16.46846528
Au 20.20010331 14.98105691 16.54420448
```

```
Au 23.07660025 14.98033373 16.57845073
Au 25.95760163 14.98143411 16.57256539
Au 28.83904836 14.98123553 16.57842672
Au 10.09438769 17.47407611 16.58761380
Au 12.97603086 17.47531535 16.59524510
Au 15.85802787 17.47836507 16.58290584
Au 18.74945611 17.48044211 16.57558569
Au 21.63285501 17.47955999 16.57933585
Au 24.51581304 17.47910197 16.57268965
Au 27.39950708 17.47915325 16.57244846
Au 30.28198964 17.47883050 16.57878835
Br 18.40341453 6.21211407 19.82818315
Br 17.78980854 13.93514900 19.50785849
Br 11.13043482 5.87925123 19.48196732
Br 10.51054917 13.69289429 19.80957089
C 11.90102752 12.43715608 20.26062781
C 14.25542383 11.76042713 20.11268036
C 13.93371483 10.48151740 20.65423119
C 11.55140378 11.20822846 20.84844950
C 14.96917140 14.48628399 19.62157733
C 12.62414615 10.28922790 21.18013700
C 12.30899979 9.18560907 22.03131211
C 11.01517671 8.91063195 22.40311114
C 9.94953073 9.72093098 21.93123474
C 15.59174102 12.12647863 19.74415602
C 15.93428581 13.45115726 19.62253867
C 13.24080094 12.79718950 20.02247798
C 10.21263380 10.84588146 21.18844738
C 18.88438592 10.14532542 22.04759368
C 16.99312368 7.44074799 20.29093675
C 17.80321830 10.94924040 22.49422234
C 16.52116511 10.67240913 22.08500043
C 16.23310187 9.57171605 21.22070131
C 18.64669632 9.02712737 21.28633044
C 17.31941466 8.66296065 20.90552773
C 14.93612738 9.37390730 20.66570567
C 14.63339923 8.09394262 20.11655411
C 13.64376722 14.13985084 19.74862343
C 13.30453463 7.71381399 19.73475053
C 12.97928629 6.38541290 19.60892461
C 13.95639628 5.36194424 19.60500681
C 15.27681816 5.72408146 19.74245029
C 15.66054366 7.06904599 20.03099356
H 9.39998958 11.49695708 20.87033138
H 8.92005240 9.46699976 22.18577190
H 10.80425834 8.05963235 23.05152516
H 13.12323465 8.54687793 22.37272559
H 16.36076231 11.36609706 19.66929474
H 16.04749030 4.95669886 19.69530826
H 13.66931415 4.31820578 19.49074022
H 12.52666921 8.46476987 19.65900100
H 19.47031620 8.38128959 20.98620856
H 15.69551749 11.30547012 22.40904588
H 17.99301374 11.79617859 23.15437385
H 19.90505222 10.39884099 22.33586028
H 15.26958183 15.52686136 19.51296238
H 12.88300231 14.91701546 19.70169129
```